# A News Recommender System Considering Temporal Dynamics and Diversity[*]

Shaina Raza

Department of Computer Science,
Ryerson University,
Toronto, ON, Canada
shaina.raza@ryerson.ca

## ABSTRACT

In a news recommender system, a reader's preferences change over time. Some preferences drift quite abruptly (short-term preferences), while others change over a longer period of time (long-term preferences). Although the existing news recommender systems consider the reader's full history, they often ignore the dynamics in the reader's behavior. Thus, they cannot meet the demand of the news readers for their time-varying preferences. In addition, the state-of-the-art news recommendation models are often focused on providing accurate predictions, which can work well in traditional recommendation scenarios. However, in a news recommender system, diversity is essential, not only to keep news readers engaged, but also to play a key role in a democratic society. In this PhD dissertation, our goal is to build a news recommender system to address these two challenges. Our system should be able to: (i) accommodate the dynamics in reader behavior; and (ii) consider both accuracy and diversity in the design of the recommendation model. Our news recommender system can also work for unprofiled, anonymous and short-term readers, by leveraging the rich side information of the news items and by including the implicit feedback in our model. We evaluate our model with multiple evaluation measures (both accuracy and diversity-oriented metrics) to demonstrate the effectiveness of our methods.

## 1 INTRODUCTION

With the advancement in internet technology, more and more news readers start to read news online. The news readers' preferences are not fixed and tend to change over time [1]. A reader's interests may be triggered by certain events or contexts [14], for example, breaking news, weather alert, a political riot in the region and such. A reader's reading behavior is also affected by temporal factors such as the time of a day or the day of a week. For example, the highest news consumption usually happens in the evening from 6-8pm, followed by a peak at 7am or sometimes around the noon [15]. We refer to these types of dynamic interests as a user's short-term preferences. A news reader may also have long-term preferences. For example, a user who is a fan of Raptors (a basketball team) may have read many basketball news about this team consistently in the past. These long-term preferences are stable and may evolve over a long period of time, possibly due to the change in the user's personality or habits or the outside environment. Besides being long or short-term, reader preferences show sequential dependencies. For example, a news reader reads about 'COVID-19 emergency' at time $t_1$, then his next action may be reading about 'COVID-19 recoveries' at $t_2$, followed by 'Stock markets after pandemic' at $t_3$. In this scenario, each of the reader's next actions sequentially depends on the prior ones.

The news recommender systems (NRS) are developed to relieve the information overload problem and suggest news items that the readers might be interested in. The traditional NRS may fail to capture the temporal dynamics (the phenomenon in which a user's preferences drift over time) in readers' preferences. Since these systems typically provide the recommendation according to the reader's full reading history, they may lose focus on the short-term preferences. Also, these systems often emphasize on high prediction accuracy. As a result, they are typically evaluated using accuracy metrics such as RMSE (root mean square error). High accuracy in a recommender system is generally associated with high personalization [9], which is no doubt important. However, in an NRS, matching everything according to readers' preferences may have adverse effects. A news reader may only be exposed to the content that conforms to his existing preferences as a consequence. This will create an echo-chamber [9] around the user where he gets all the information echoing what he already thinks. At the society level, this will create a filter bubble around the like-minded news readers and they tend to isolate themselves from different viewpoints and perspectives [9]. For example, the readers in a filter bubble might only get news items that are aligned with their political beliefs, not the ones that contradict with them. Adding diversity can solve this problem to a certain degree.

There always exists a trade-off between the accuracy and the diversity. Currently in the literature, the NRS could go beyond accuracy by expanding the reader's current list of recommendations to include some diverse or novel items. In this re-ranking process to include diversity, the dynamics in readers' preferences are ignored. To address this issue, we need to include diversity, just like accuracy, in the design phase of an NRS. We also need to have the right balance between the accuracy and the diversity in the NRS.

A few recent NRS [1,23] have started to focus on timeliness to address the temporal dynamics issue. However, these models suffer from the data sparsity problem because of the limited reader interactions on the news platform. There are also anonymous or short-term users without any explicit profiling information. These readers may click a few or no news items displayed in a session. A possible solution is to exploit the rich information from items' metadata such as titles [19]. For example, the user representations are learnt from the browsed and matching candidate news titles. However, relying on a single piece of information such as titles may not be enough to learn accurate news and user representations. There are other pieces of information that may be more descriptive (e.g., the text of a news story) or that are better indicators of the reader's short-term or long-term preferences (e.g., topics, categories). Also, these models are accuracy-centric (built to improve the recommendation accuracy), without considering diversity in the design of NRS.

Motivated by the above line of reasoning, we aim to find solutions that can include the temporal dynamics in an NRS and in the meantime model diversity as a key factor in the design stage. In this PhD research, we have the following long-term goals: (i) temporal dynamics in readers' preferences; (ii) diversity as the key design principle; and (iii) data sparsity problem. Besides these long-term goals, we also have these short-term objectives: (i) to identify the sequential patterns in the reader's consumption history, along with short-term and long-term reader preferences, in order to have a more complete view of temporal dynamics; (ii) to exploit the rich side information from the news items to learn more accurate item and user representations; and (iii) to include diversity in the design and optimization phase of the recommendation model.

The rest of the paper is organized as follows: Section 2 is about the related work. Section 3 discusses about our proposed approach. Section 4 briefly summarizes some of the results we have obtained so far. Finally, Section 5 gives the conclusion and briefly discusses the future work.

## 2 RELATED WORK

### 2.1 Temporal dynamics

Temporal dynamics in recommender systems refer to the concept of accurately capturing user preferences over time. The earlier model uses simple time-decay functions [6]. It favours users' short-term preferences, and the older data is often discarded. To determine the long-term preferences, the full user history is needed. This can be accomplished by models such as timeSVD++ [11], in which the timeline of the user's preferences is split into static time bins. This model favours users' long-term preferences. There is another model [20] that combines both the long-term and short-term user preferences, however, time-decay is not considered in terms of item popularity and drifting user interests. There are also the graph-based methods, where users, items and the time factors are included as nodes, thus making a tri-partite graph [21]. An issue with graph-based models is that they rely heavily on content-based features that are not always available to the system and could be quite costly to collect. The temporal models listed above are for general recommender systems.

There is limited work on NRS that addresses the temporal dynamics issues. CLEF NewsREEL [12] offered a campaign-style evaluation lab where the goal was to make news recommendation that must respond to a request within a given time frame (100ms). It emphasizes on the timeliness of the news recommendation. However, it is not capturing the dynamics part in readers' preferences. A general limitation of these traditional methods is that they all rely on hand-crafted feature engineering to model users' preferences and they treat time as the static feature of user preferences.

The above-mentioned methods do not consider the sequential dependencies among the user interactions. A sequential recommendation task takes into account users' short-term interests with all interactions happening within a close proximity of time (i.e., in a session). Thus, news reading activity can be considered as a sequential task. The Markov model is a well-known sequence modeling technique, which has been used in NRS [3]. The Markov model represents sequential data as a stochastic process over discrete random states (variables) where every state depends on the previous state. A limitation of this model is that when including higher states, the computational complexity of the model is drastically increased. Another sequential modeling approach used in NRS is the Reinforcement Learning (RL) model with actions and rewards [23], where the model dynamically adapts recommendations to future rewards. This model also suffers from the computational complexity problem brought by the growing states. Due to the importance of the sequential modeling to a recommender system, reducing complexity of these models is an ongoing research topic.

The Recurrent Neural Network (RNN) and its variants Long short-term memory (LSTM) and Gated recurrent unit (GRU) are well-known models to handle the sequential patterns in user behaviors for the session-based recommendation tasks. RNNs are used in a few NRS to learn user representations (patterns in a user's browsing history) [1,16,22]. These methods have hidden states unlike Markov and RL methods, they are faster and they are more robust. On top of temporal factors, it is also possible to include other feasible contextual factors as the intermediary layers in an RNN model. This is yet to be explored, but it could be useful in an NRS with news and user related side information used as contextual factors.

Lately, the transformer-based methods [18] that use attention mechanisms are being widely adopted for sequence-to-sequence (Seq2Seq) tasks. The well-known Bidirectional Encoder Representations from Transformers (BERT) could extract the sequential relations from the text (sentences). However, this is not equivalent to including the time-ordered user interactions in the model. Attempts have been made to apply BERT on a recommender system as the downstream task [17]. In this model [17], a sequence of readers' interactions that are chronologically ordered are being created to feed into the transformer model to make recommendations. These models do not have a separate user model, which could be essential for temporal dynamics. There is also no consideration of the explicit user or item side information.

## 2.2 Diversity

Recommendation is a two-step process, i.e., the rating prediction and the top-N item recommendation based on predicted ratings. Diversity is usually incorporated into the second step as re-ranking of recommendations. Re-ranking is an expensive process since recommendations are made for each user independently. There is very limited work that considers the diversity in the design phase of the recommendation models. For example, the variance-minimization penalty is used in a diversity-oriented recommender system [8] to reduce the bias in the predicted ratings towards particular item groups. In another recommender system [2], the minimum-cost network flow with subgraphs is used to include diversity in the recommendation process. An issue with these models is that they are good for small-scaled synthetic data and may suffer from data sparsity problem when dealing with the high-dimensional news datasets. There is also limited work in DL-based recommenders [23] that considers diversity. However, when we ran these models, we observe that the output from these DL models are quite similar (not diverse) at different time steps when the training samples at each time step are the same.

The recommender systems mostly adopt the Tikhonov regularization in the learning algorithm to achieve accuracy in approximate solutions [11]. The literature shows that Tikhonov regularization with $l_2$-norm (or penalty) is a desirable approach when we aim for high prediction accuracy [7]. However, it is not optimal for achieving diversity. There are other diversity-promoting regularization terms, but this is rather an unexplored area in the state-of-the-art.

# 3 PROPOSED APPROACH

We have identified the research gaps in the previous section. In this section, we define our research directions and briefly explain how we address these issues in our dissertation work. First, we conduct an in-depth study of the NRS and then we put more focus on the special features of news domain as discussed below.

## 3.1 Temporal dynamics in NRS

We aim to propose novel mechanisms to address the temporal dynamics issue in an NRS by exploring news readers' long-term, short-term and sequential reading preferences. Based on our extensive literature review, we find that the temporal dynamics in user behavior is not just a time-dependent activity but there are many other factors that influence it. The side information associated with either users or items plays a crucial role in this regard. The title, text body, category, subcategory, topics are all useful for representing news. Different type of side information contains different level of details, e.g., the full text of a news story is more descriptive than the title alone. We also find that news articles accessed by a reader during different time windows convey different types of information regarding his reading patterns. For example, if we analyze the full interaction history of a news reader, we may find the general likings of the reader on news such as Sports, Politics, Entertainment. If we analyze the interaction history in a session, we may find different sets of topics or words that reflect his sequential and short-term interests.

We plan to perform this part of the work using two types of algorithms: first using a computationally less expensive and more interpretable approach such as matrix factorization (MF) [15], and then using more powerful and robust DL approaches for sequential modeling, such as RNNs (LSTM and GRU), attention-mechanisms. The goal is to test the efficacy of using different algorithms under a certain experimental setup. We also adopt a time-aware evaluation strategy. This includes a time-series analysis on the data (before the experiments) and the time-based split of the data for the train and test sets (after tuning all hyper-parameters).

Based on our time-series analysis, we find that different temporal resolutions when being added as the side information with the full timestamp of the user's rating help us differentiate the long-lasting reader preferences from the transient ones. The coarser-level time units (year, month, day) reflect a reader's long-term preference, whereas the finer-grained time units (hour, minute, second) help us understand the reader's short-term preferences. We also find that if we simply look at the user ratings on individual news items, it might not be easy to get a clear view of their general interests. If we link these ratings to the side information of the news items such as sources, topics, or categories, it would be easier to identify the pattern. The long-term user interests can be utilized to identify the type (category) of news that users are interested in, and then the short-term interests help the NRS filter specific news items for a reader at that time. This part of the work is implemented using MF methods.

During our experiments, we have made two observations. First, explicit user feedback is scarce, and second, it is difficult to differentiate the positive implicit feedback signals (e.g., clicks, scrolls, time spent) from the negative ones. To solve the sparse explicit feedback problem, we may consider the implicit signals, where we treat everything a reader interacts with as a positive feedback and everything a reader does not interact with as a negative feedback. The assumption is that an interaction is an indicator that the user likes the news item. However, some of the interactions in the implicit feedback could be the result of the reader's boredom, random clicks or idle time. Also, some of the non-interactions could be the result of the reader's unawareness of those news items. Although currently we take a simplified approach, treating all interactions as positive implicit feedbacks, we still explore other options to better deal with this issue.

In order to train our model, we need the negative samples besides the positive samples for the implicit reader feedback. For that, we randomly generate sample news items from the same session which are not clicked by this reader [10]. We implement a DL-based model where we learn both the readers' long-term and short-term preferences. We learn readers' long-term preferences using their whole consumption history. In order to learn the readers' short-term preferences, we consider their recently browsed history through the RNN models. We also consider the rich side information from the news items to learn news representations. We make use of attention mechanism [18] to learn useful information from both the news and readers' representations.

## 3.2 Diversity in NRS

In our work, we include both the accuracy and the diversity in the optimization phase. We again adopt two different algorithms here: the computationally inexpensive and linear approach using the latent factor model and the DL-based approach. By default, a recommender system is built to reduce the sum of squared errors in the known ratings and the predicted ratings. The Tikhonov regularization with $l_2$-norm (Ridge regression) is considered as de-facto regularization in this regard. It gives us the approximate solution to produce a lower (not exactly zero) mean squared error in predictions. However, we may utilize other forms of regularizations such as $l_1$-norm that chooses the highly correlated features from the data and gives more absolute solutions (close to zero). In this way, we sacrifice some level of accuracy at the cost of diversity. We can also control the relative balance and the intensity of these regularizations to get a trade-off. Although the usage of these regularizations is not novel, no previous work has applied both regularization terms to address the accuracy and diversity in a unified optimization framework. In our latent factor model, we test different form of regularization terms from the Ridge, Lasso and Elastic-net regressions. The goal is to include both the accuracy term and the diversity term in our model. In our work, we also use the reweighted optimization algorithms to optimize the model on different form of penalties ($l_\alpha$-norm) from these regularizations.

In our DL-based approach, our work is motivated by the observation that reader interests in a news domain are quite volatile and diverse and are triggered by specific contexts and temporal factors. Some of reader interests may last for a longer time and are quite constant for the same user. However, if we check the reader preference in a session, we may find that it could be changing quite abruptly. Based on this observation, we use RNN models to reflect diversity in the short-term preference model.

## 3.3 Multi-criteria evaluation

Normally, the machine learning models are evaluated using RMSE to check for the prediction accuracy. However, in a recommender system, we also need to see if the user is getting the relevant (good) recommendations. So, we need to evaluate our models using other measures such as precision @k (the proportion of the top k results that are relevant) and recall @k (the proportion of all relevant results included in the top k). We also take the F1-score @k (harmonic mean of both) as one metric. We take RMSE, precision, recall and F1-score as our accuracy metrics.

Since diversity is a key principle in our design, we also need to evaluate our model on diversity and novelty. In our work, we take both (diversity and novelty) as our diversity metrics. We compute the diversity as the average dissimilarity of all pairs of items in a user's recommended list. We compute the novelty as the ratio of how many recommended items are unknown to a user.

Recommendations high on accuracy metrics are more biased towards popular items or items that are rated extensively by users. The diverse recommendations, on the other hand, are less biased towards popular items and less aligned with users' preferences. Both diversity and novelty trade off recommendation accuracy but the difference between these two is that the former recommends less biased items from the seen list of recommended items, whereas the latter makes recommendations from the unseen items. Since our primary goal is to include both diversity and accuracy in a unified optimization framework, we define a composite metric to measure the trade-off between these measures, which is the weighted sum of the F1-score and the combined diversity scores (mean of diversity and novelty).

## 4 PRELIMINARY RESULTS

In this section, we briefly describe our models first and then we summarize the results we have obtained so far for the research goals described above.

In our MF model [15], we consider the reader's time of rating and the taxonomy (category-subcategory) of news items as the baseline predictors. A baseline predictor represents the effects associated with the biases (effect of one entity on the other) without involving user-item interactions [11]. The goal of this model is to predict the ratings of a news reader for the unrated news items and to make recommendations based on predicted ratings. We consider different combinations of time-varying factors to integrate the temporal dynamics. In that, we first include the main timestamp associated with the reader's rating as the baseline predictor, and then we supplement the varying time unit (year, month, day, hour, minute, second) along with the whole timestamp. The point is to capture either the rapidly changing effects (through fine-grained timing information as biases) or the steady temporal effects (through coarser time units as biases) in readers' behavior. We also add the category and the subcategory information in the baseline predictors. The category information is item-related, and it also reflects a news reader's long-term preferences. In this model, the timestamp of rating, the time units, the category and the subcategory are considered as side information.

Our second approach to implement temporal dynamics is a DL-based approach. We make use of the Transformer architecture [4] in this model for two prominent tasks: to capture the sequential reading behavior of a news reader and to perform the next click prediction. We build separate reader and news components in our proposed framework. Our reader component consists of two modules: the long-term preference module and the short-term preference module. In the long-term module, we consider the reader's full consumption history. In the short-term module, we apply the LSTMs on the current session in which the reader has interacted with news items. We also learn rich news representations in our news component by including the side information from the headline, category, subcategory and snippet (body) of each news item. We use the attention mechanism [18] to learn important information from each piece of side information in the news component. The news component is then made available for the reader component to learn reader representations.

To include diversity in NRS, we also consider two approaches: a simple latent-factor model and the DL-based approach. We encode the reader information and their past interactions in our latent factor model. The typical latent factor models in the recommender system [11] generally suffer from the scalability issue. Also, these models tend to favor popular items (no diversity). Therefore, we use Generalized Linear Model (GLM) (regression-based methods) to factorize a large user-item rating matrix and include diversity in our model through different regularization terms (other than Tikhonov). GLM is found to be mathematically suitable for larger datasets that uses probabilistic modeling to represent uncertainty in ratings [5]. It also allows us to include rich user-item interactions and regularize the model for both accuracy and diversity.

In our second approach to diversity, we consider the short-term preference model in our DL-based approach. We use the RNNs, specifically, LSTM to learn short-term reader representations from their recent browsing histories. We use the word-level attention in the news component. Since our reader component depends on the news component, the same attention-mechanism is also used by our short-term preference module to learn diverse but important reader representations. We represent the short-term reader representation in the final hidden state of the LSTM network.

The findings from the results of running these models are given below. First, we observe that including the temporal information in all the models gives us better results on prediction accuracy (lower RMSE) and recommendation accuracy (higher F1-score). In our first approach using MF methods, as we continue to include more side information such as categories and subcategories, the prediction accuracy and recommendation results get better, compared to those with partial or no side information. We also find that including more fine-grained timing information as the side information in our MF model further improves the results. In terms of the recommendation results, our MF model outperforms the DL-based baselines for F1-score @10 and 20, i.e., when the k value is small. However, as we increase the value of k, the DL-based recommenders show better accuracy. Typically, in a real recommendation scenario, readers only have the patience to check the first 10 or 20 recommended results. So, our model works better in typical cases. Overall, our MF-based model is simpler, takes much less training time and is comparable to (sometimes even better than) the contemporary computationally complex DL-based recommenders.

In our approach to include diversity as the design principle in the recommendation model, we observe that there is a negative correlation between the accuracy and the diversity of the recommendation results. When we optimize our model for higher accuracy, the diversity drops, and vice versa. We also find that the better prediction accuracy (lower RMSE) of a model does not necessarily guarantee the better recommendation accuracy (higher F1-score). Another observation from our results is that as we include more dataset features to train the model, the accuracy of the model gets better. The same observation can be made in both MF and DL-based approaches. However, the DL-based approach gives us the best results in terms of model accuracy and the recommendation results, because they have higher capacity to easily integrate the side information.

In our latent factor model, we get the best results when we include both the accuracy and the diversity in our NRS. This is also validated by the better trade-off score (composite metric combining accuracy and diversity) of our model compared to others. By including both the accuracy and diversity terms within a single optimization framework, we can get better results than the related baselines such as SLIM model [13] and the state-of-the-art DL-based recommenders. A few baselines might have obtained better diversity scores at some points, but if we look at the overall results, our models provide the right balance.

In our DL-based approach, we find that when we include both the short-term and long-term user preferences, we get better results compared to the case when we include only one of them. Our results show that the short-term preference module plays an important role in the diversity score of our model. Another observation we get from our results is that the attention mechanism helps us select important words within the context of each news item and its side information. As we increase the number of heads in our model, the results get better, however, with more layers and more heads, the computational performance of the model degrades, and the precision of the results drops.

The results also show that an NRS works better if we consider the implicit reader feedback (i.e., everything that a reader clicks, whether he likes it or not). This is demonstrated by the results when comparing among our own approaches. We could also deal with the data sparsity problem when including implicit feedbacks. The explicit feedback is derived from the readers' comments (sentiment scores) on news items and is used in our MF and latent-factor GLM-based approaches. We consider the implicit feedback in our DL-based approach. A preliminary test on including both the explicit and implicit feedbacks showed us that results with implicit feedbacks are better than those without them. This is because in an NRS, a news reader may still read about a topic or a news story which he has not liked in the past, just for gaining knowledge, getting information, or learning the latest update on an event. Although the implicit feedback may bring some inaccuracy into the model as we discussed earlier, it is kind of aligned with the real-world scenario where people consume different news articles even if they don't like them. Overall, our DL-based model with more side information included in the news component and with the combination of both the long-term and short-term components in the reader model achieves the best results.

## 5 CONCLUSION

In this paper we present the current status of a dissertation aimed at developing a news recommender system. Among the challenges of the news domain, we aim to address the most pressing ones (temporal dynamics, diversity and the data sparsity) in our work. The goal is to enhance the reader experience through the use of the NRS. We first address the temporal dynamics issue in user preferences. We learn both the long-term and short-term preferences of news readers. We also learn the sequential patterns from the reader's consumption history. Next, we include the diversity as the key design principle in the NRS. We propose to include different forms of regularization terms during the optimization phase to trade off accuracy with diversity. We find that it is important to retain the accuracy with the requisite level of diversity. For both the temporal dynamics and the diversity in the NRS, we adopt two different approaches: computationally simple, interpretable and linear models, and the DL-based model. We also evaluate our system using different criteria to check the results of our models from different perspectives.

Although we have conducted a preliminary test on our models and some of the results are validated through publication, we believe that we need more research to explore further. Specifically, we would like to know whether and how we should have more detailed considerations (e.g., a threshold for the level of accuracy and diversity) to address the accuracy-diversity dilemma in an NRS. We would like to investigate more on how to include diversity in the design of the recommendation model in our DL-based approach. We need to explore more on the issue of missing and negative implicit feedbacks. We also want our NRS to be a democratic recommender. For that, we need to consider other aspects such as assimilating bias, neutralizing polarization, or breaking filter bubbles. We also need to study various post-algorithmic effects from news recommenders on reader's behavior.

### ACKNOWLEDGEMENT
Special thanks to my advisor Dr. Chen Ding for her valuable support and guidance.